\documentclass[11pt]{article}

\pdfoutput=1
 
\usepackage{amsfonts,amssymb,amsmath,amscd,amsthm,latexsym}

\usepackage{graphicx}
\usepackage{hyperref}
\usepackage{natbib}
\usepackage{fullpage}
\usepackage{setspace} 
\theoremstyle{plain}
\newtheorem{theoremeeng}{Theorem}[section]
\newtheorem{lemmeeng}{Lemma}[section]
\newtheorem{proposition}{Proposition}[section]
\newtheorem{definition}{Definition}[section]

\newtheorem{algo}{Algorithm}

\newcommand{\Vect}[1]{{\bf {#1}}}
\newcommand{\bX}{\Vect{X}}
\newcommand{\bY}{\Vect{Y}}
\newcommand{\by}{\Vect{y}}
\newcommand{\bZ}{\Vect{Z}}
\newcommand{\bx}{\Vect{x}}
\newcommand{\bz}{\Vect{z}}

\newcommand{\Scal}[3]{<#2,#3>_{#1}}
\newcommand{\Input}{E}

\newcommand{\Preuve}{\noindent\textsc{\bf Proof}}
\newcommand{\Fin}{$\Box$\\}	
\def \I {\mathbb{I}}

\usepackage{geometry}
\title{Maximin design on non hypercube domain and kernel interpolation}

\author{
Yves Auffray \\
Dassault Aviation \& \\
D\'epartement de Math\'ematiques, \\ 
Universit\'e Paris-Sud, France
\and	
Pierre Barbillon \\
INRIA Saclay, projet \textsc{select}, \\
D\'epartement de Math\'ematiques, \\ 
Universit\'e Paris-Sud, France
\and
Jean-Michel Marin
\footnote{Corresponding author: place Eug\`ene Bataillon, Case Courrier 051, 34095 Montpellier cedex 5}
\footnote{\textsc{jean-michel.marin@univ-montp2.fr}} \\
Institut de Math\'ematiques et Mod\'elisation de Montpellier \\
Universit\'e Montpellier 2
}

\begin{document}

\maketitle

\begin{abstract}

In the paradigm of computer experiments, the choice of an experimental design is an important issue.
When no information is available about the black-box function to be approximated, an exploratory design has to be used.
In this context, two dispersion criteria are usually considered: the \textsc{minimax}
and the \textsc{maximin} ones. In the case of a hypercube domain, a standard strategy consists
of taking the \textsc{maximin} design within the class of Latin hypercube designs.
However, in a non hypercube context, it does not make sense to use the Latin hypercube strategy.
Moreover, whatever the design is, the black-box function is typically approximated thanks to kernel
interpolation. Here, we first provide a theoretical justification to the \textsc{maximin} criterion with respect to kernel interpolations.
Then, we propose simulated annealing algorithms to determine \textsc{maximin} designs in any bounded connected domain.
We prove the convergence of the different schemes. Finally, the methodology is applied on a challenging real example
where the black-blox function describes the behaviour of an aircraft engine.\\

\vspace{0.2cm} \noindent {\bf Keywords:} 
computer experiments,
kernel interpolation, 
Kriging, 
\textsc{maximin} designs, 
simulated annealing.

\end{abstract}


\section{Introduction}

A function $f:E\rightarrow \mathbb{R}$ is said to be an expensive black-box function if $f$ is only known through
a time consuming code. It is assumed that $E$ is enclosed in a known bounded set of $\mathbb{R}^d$.
$E$ is not necessarily a hypercube domain or explicit. $E$ may be given through an indicator function only.
It is assumed that testing if a point belongs to $E$ is not burdensome. Hence simulate a point in $E$ can be performed
by sampling rejection.
In order to deal with some concerns such as pre-visualization, prediction, optimization and probabilistic analysis
which depend on $f$, an approximation of $f$ is usually used. This is the paradigm of computer experiments
\citep{santner:williams:notz:2003,fang:li:sudjianto:2005} where the unknown function $f$ is deterministic.
An approximation of $f$ can be obtained using a kernel interpolation method \citep{schaback:1995,schaback:2007}.
For the corresponding covariance structure, the Best Linear Unbiased Predictor (BLUP) given by Kriging metamodeling
\citep{matheron:1963} provides the same approximation of $f$.
Due to its flexibility and its adaptivity to non-linear functions,
Kriging is one of the most used approximation method by the computer experiments' community.
In our tests, we have observed that Kriging works well for approximating a large class
of non linear functions when the input space is of dimension less than ten.
For more details on Kriging, one can see for instance:
\cite{sacks:etal:1989:1,sacks:etal:1989:2,cressie:1993,laslett:1994,stein:1999,stein:2002,li:sudjianto:2005,joseph:2006,denhertog:kleijnen:siem:2006}. 
 
\vspace{0.5cm} The kernel interpolation methodology needs the choice of a kernel $K$ (kernel satisfying some conditions detailed below)
and a design $\bX=\{\bx_1,\ldots\bx_N\}$ where the function $f$ is to be evaluated, giving $\{f(\bx_1),\ldots,f(\bx_N)\}$.
As it is well-known, a space of functions $\mathcal{H}_K$ is associated to $K$.
If the function $f$ belongs to $\mathcal{H}_K$, we can control the pointwise error of $s_{K,\bX}(f)$, the interpolator
of $f$ on $\bX$. In this deterministic paradigm (the function $f$ is not random), there are essentially
two main kinds of properties that a design can have \citep{koehler:owen:1996}:
\begin{itemize}
\item projection properties such as Latin hypercube designs \citep{mckay:beckman:conover:1979} or its generalization
Latin hyper-rectangle sampling which allows for non-equal cell probabilities \citep{mease:bingham:2006};
\item exploratory properties which are warranted by criteria such as:
\begin{itemize}
\item \textsc{minimax} which means that the design has to minimize
 \begin{equation}
 \label{minimaxmd}
h_{\bX}=\sup_{\by\in E}\min_{1\le i \le N}\Vert \by-\bx_i\Vert , 
\end{equation}
\item \textsc{maximin} which means that the design has to maximize
\begin{equation}
 \label{maximinmd}
\delta_{\bX}=\min_{1\le i,j\le N}\Vert \bx_i-\bx_j\Vert\,.
\end{equation}
Moreover, between two designs $\bX_1$ and $\bX_2$ such that
$\delta_{\bX_1}=\delta_{\bX_2}$, using the \textsc{maximin} criterion, we choose the design for which the number
of pairs of points with distance equal to $\delta_{\bX_1}$ is minimal.
\item Mean Squared Error (MSE) based criteria.
These criteria are linked to the MSE of the BLUP in the context of Kriging metamodeling. Designs can be 
sought to minimize the Integrated Mean Squared Error (IMSE) of the BLUP over the domain $E$ or to minimize the maximum
of the MSE over $E$.
\end{itemize}
\end{itemize}
The \textsc{minimax} and \textsc{maximin} criteria have been proposed for Kriging metamodeling by 
\citet{johnson:etal:1990}. \citet{sacks:etal:1989:2} have detailed the MSE-based criteria.
For others criteria, one can see \citep{bursztyn:steinberg:2006}.

For kernels defined by radial basis functions, \citet{schaback:1995} and \citet{madych:nelson:1992} have shown that
the \textsc{mininax} criterion $h_{\bX}$ explicitly intervenes in an upper bound on the pointwise error between $f$ and
$s_{K,\bX}(f)$. The upper bound has the form $G(h_{\bX})$ where $G$ is an increasing function $\mathbb{R}_+\rightarrow \mathbb{R}_+$.
Here, we prove that $h_{\bX}\le \delta_{\bX}$ and then, that a \textsc{maximin} design also provides an uniform upper bound
of the pointwise error.

\textsc{minimax} and IMSE criteria are costly to evaluate and, typically, the \textsc{maximin} criterion is privileged.
In the case where $E$ is a hypercubic set, \citet{morris:mitchell:1995} provided an algorithm based on simulated annealing
to obtain a design very close to a \textsc{maximin} Latin hypercube designs, (the criterion optimized is not exactly the
\textsc{maximin} one). For the two-dimensional case, \cite{vandam:husslage:denhertog:melissen:2007}
derived explicit constructions for \textsc{maximin} Latin hypercube designs when the distance measure is $L_\infty$
or $L_1$. For the $L_2$ distance measure, using a branch-and-bound algorithm,
 they obtained \textsc{maximin} Latin hypercube designs for $N\leq70$.

For some non hypercubic domains, the use of projection properties can lead to poor exploratory designs.
For instance, if $E=\{\left(x^1,x^2\right)\in[0,1]^2:x^1\geq x^2\}$ then, the only Latin hypercube design is 
on the line $x^1=x^2$. Moreover, in some cases, they are impossible to satisfy. Therefore, we focus on exploratory properties only.

In the case of an explicit constrained subset of $[0,1]^d$, \cite{stinstra:denhertog:stehouwer:vetjens:2003} proposed an algorithm
based on the use of NLP solvers. Here, we propose some algorithms to achieve a \textsc{maximin} design for general (even not explicit)
non hypercubic domains. Our schemes are based on simulated annealing. Our proposals are not heuristic, we study the convergence properties
of the schemes proposed. 

Recall that the simulated annealing algorithm aims at finding a global extremum of a function by using a Markovian kernel which
is the composition of an exploratory kernel and an acceptance step depending on a temperature which decreases during the iterations. \\
In some presentations (e.g. \cite{bartoli:delmoral:2001}), the simulated annealing algorithm 
is based on a Metropolis-Hastings algorithm \citep{chib:1995}.
In that case, it can be proved that the resulting Markov chain tends to concentrate
on a global extremum of the function to be optimized with high probability 
when the number of iterations tends to infinity.
Here, we introduce a simulated annealing scheme based on a Metropolis-within-Gibbs algorithm
\citep{roberts:rosenthal:2006} and prove its convergence.

The paper is organized as follows, in Section 2 the kernel interpolation method is described and a theoretical justification of the
\textsc{minimax} and \textsc{maximin} criteria is provided thanks to the pointwise error bound between the interpolator and the function $f$.
Then, in Section 3 the simulated annealing algorithm is presented. A proof of convergence is given. Section 4 deals with the case where $E$ is
not explicit and can only be known by an indicator function. Two variants of the algorithm are proposed and their theoretical properties are stated.
In Section 5, the algorithms are tried on some examples and practical issues are discussed. Finally, in a last Section, the methodology is
applied on a real example for which the domain is not an hypercube.

\section{Error bounds with kernel interpolations}

A kernel is a symmetric function $K:E\times E\rightarrow \mathbb{R}$ where $E$ is the input space which is assumed to be bounded.
The kernel has to be at least conditionally positive definite to be used in kernel interpolation. For the sake of simplicity,
kernel interpolation is presented for positive definite kernels only.
$\mathbb{R}^E$ denotes the space of functions from $E$ to $\mathbb{R}$.

\begin{definition}
A kernel $K$ is positive definite if $$\forall (\zeta_1,\bx_1)\ldots (\zeta_N,\bx_N)\in\mathbb{R}
\times \Input, \sum_{1\le l,m\le N}\zeta_l\zeta_mK(\bx_l,\bx_m)\ge 0\,.$$
\end{definition}
For any $\bx\in\Input$, let $K_{\bx}$ denote the partial function $\bx'\in\Input\mapsto K(\bx,\bx')\in \mathbb{R}$. The linear combinations of
functions taken in $\{K_{\bx},\bx\in\Input\}$ span a functional pre-Hilbert space $\mathcal{F}_K$ where 
$$
\Scal{\mathcal{F}_K}{\sum_{l=1}^L\zeta_lK_{\bx_l}}{\sum_{m=1}^M \mu_mK_{\bx'_m}} = \sum_{m=1}^M\sum_{l=1}^L\zeta_l\mu_m K(\bx_l,\bx'_m)
$$
is the scalar product. Aronszajn's theorem states that there exists a unique space $\mathcal{H}_K$ which is a completion of
$\mathcal{F}_K$ where the following reproducing property holds
\begin{equation*}
\label{reprorkhsmd}
\forall f\in\mathcal{H}_{K},\ \bx\in\Input,\ f(\bx)= \Scal{\mathcal{H}_{K}}{f}{K_{\bx}}\,.
\end{equation*}
$\mathcal{H}_K$ is called a Reproducing Kernel Hilbert Space (RKHS). \\
Let us denote by $s_{K,\bX}(f)$ the orthogonal projection of $f$ on $\mathcal{H}_K(\bX)=\mbox{span}\{K_{\bx_1},\ldots,K_{\bx_N}\}$
($f$ is assumed to be in $\mathcal{H}_K$; $\bX=\{\bx_1,\ldots,\bx_N\}$ and $K$ are given).

\newpage

\begin{lemmeeng}
$s_{K,\bX}(f)$ interpolates $f$ on $\bX$.
Among the interpolators of $f$ on $\bX$, $s_{K,\bX}(f)$ has the smallest norm: $s_{K,\bX}(f)$ is the solution of the following problem
\begin{center}
$\left\{
\begin{array}{l}
\min_{g\in {\cal H}_K}\Vert g \Vert_{{\cal H}_K}\\
g(\bx_k)=f(\bx_k), \ k=1,\dots N
\end{array}
\right. .$
\end{center}
\end{lemmeeng}
\noindent This interpolator corresponds to the BLUP in the Kriging literature \citep{cressie:1993,stein:2002}.
It has a Lagrangian formulation.
\begin{lemmeeng}
For any $\bx\in E$,
 \begin{equation*}
  s_{K,\bX}(f)(\bx)=\sum_{i=1}^N u_i(\bx) f(\bx_i)
 \end{equation*}
where the functions $(u_i:E\rightarrow \mathbb{R})\in \mathcal{H}_K(\bX)$ are such that, $\forall \ 1\le i \le N$,
$$\left\{
\begin{array}{ll}
   u_i(\bx_i)=1&\\
u_i(\bx_k)=0& \text{if } k\neq i
  \end{array}\right.\,,
$$
and
\begin{equation*}
\label{eqU}
K[\bX,\bx]=K[\bX,\bX]U(\bx)\,,
\end{equation*}
where $U(\bx)=\left( \begin{array}{c} u_1(\bx)\\ \vdots \\ u_N(\bx)  \end{array}\right)$,
$K[\bX,\bx]=\left( \begin{array}{c} K(\bx_1,\bx)\\ \vdots \\ K(\bx_N,\bx) \end{array}\right)$ and $K[\bX,\bX]$ is such that \\
$(K[\bX,\bX])_{1\leq i,j\leq N}=K(\bx_i,\bx_j)$. 
\end{lemmeeng}
Hence, the pointwise error can be bounded from above, $\forall \bx\in E$
\begin{equation*}
|f(\bx)-s_{K,\bX}(f)(\bx)|=|\Scal{\mathcal{H}_K}{f}{K_{\bx} -\sum_{i=1}^N{u_i(\bx)K_{\bx_i}}}|\leq
\Vert f\Vert_{{\cal H}_K}\Vert K_{\bx}-\sum_{i=1}^N{u_i(\bx)K_{\bx_i}}\Vert_{{\cal H}_K}.
\end{equation*}

Let $P_{\bX}(\bx)=\Vert K_{\bx}-\sum_{i=1}^N{u_i(\bx)K_{\bx_i}}\Vert_{{\cal H}_K}$. $P_{\bX}$ depends only on the kernel $K$ and
on the design $\bX$. $P_{\bX}$ corresponds to MSE of the BLUP.
When it is integrated on the domain $E$, we obtain  the IMSE criterion.
As already explained, the IMSE or the the maximum MSE can be minimized
to determine an exploratory design. However, it depends on the kernel and it is costly to compute. \\
For some kernels $K$ defined by $K(\bx,\bx')=\phi(\Vert\bx-\bx'\Vert_2)$ where $\Vert\bx\Vert_2=\sqrt{\sum_{i=1}^d\left(x^i\right)^2}$,
($\bx\in E$) and $\phi:\mathbb{R}_+\rightarrow \mathbb{R}$, \citet{schaback:1995} provides the following upper bound on $P_{\bX}(\bx)$:
$$
P_{\bX}(\bx)\le G_K(h_{\bX})\,.
$$
The quantity $h_{\bX}=\sup_{\by\in E}\min_{1\le i \le N}\Vert \by-\bx_i\Vert$ is associated to the \textsc{minimax} criterion. 
$G_K$ is an increasing function, obviously depending on the kernel. The smoother the kernel $K$, the faster $G_K(h)$ tends to $0$
for $h\xrightarrow{>}0$. For instance, the Gaussian kernel is defined by $K(\bx,\bx')=e^{-\theta\Vert \bx-\bx'\Vert_2^2}$
where $\theta$ is a real positive parameter; in that case, $G_K(h)=Ce^{-\delta/ h^2}$ where $C$ and $\delta$ are constants depending on $\theta$.
By choosing a design $\bX$ with a low $h_{\bX}$, one thus ensures a small pointwise error independently of the chosen 
(radially symmetric) kernel. This justifies using \textsc{minimax} (\ref{minimaxmd}) optimal designs.
The next proposition shows that a bound on the pointwise interpolation error is still guaranteed when a 
\textsc{maximin} optimal design is used.
\begin{proposition}
If $\bX$ is a \textsc{maximin} design, $E$ is enclosed in the union of the balls of center $\bx_i$ and of radius $\delta_{\bX}=\min_{1\le i,j\le N}\Vert \bx_i-\bx_j\Vert$.
\end{proposition}

\Preuve \\
This proposition is proved by contradiction: 
let $\bX$ be a \textsc{maximin} design and let us suppose that there exists a point $\bx_0\in E$ such that
$\Vert \bx_0-\bx_i\Vert >\delta_{\bX}$ for all $\bx_i\in \bX$.

\noindent Let $(\bx_{i_0},\bx_{j_0})\in \bX^2$ be a pair of points such that $\Vert \bx_{i_0}-\bx_{j_0}\Vert = \delta_{\bX}$ and construct
the design $\bX'=\{ \bx_1\dots \bx_{i_0-1},\bx_{0}, \bx_{i_0+1}\dots \bx_N\}$ where the point $\bx_{i_0}$ is replaced by
the point $\bx_0$. \\
$\delta_{\bX'}\geq \delta_{\bX}$ and, in the case $\delta_{\bX'}=\delta_{\bX}$,
$\bX'$ is better than $\bX$ with respect to the \textsc{maximin} criterion because the $\bX'$ contains less pairs of points
for which the distance is equal to $\delta_{\bX}$.\\
Thus, there is a contradiction because $\bX$ is not a \textsc{maximin} design.
Hence, for any $\bx\in E$, there exists a $\bx_i\in\bX$ such that $\Vert \bx-\bx_i\Vert \leq \delta_{\bX}$. 
\Fin

As a consequence of this proposition, if $\bX$ is a \textsc{maximin} design,
\begin{equation*}
|f(\bx)-s_{K,\bX}(f)(\bx)|\leq
\Vert f\Vert_{{\cal H}_K} G_{K}(\delta_{\bX}).
\end{equation*}
This result justifies theoretically the use of \textsc{maximin} designs when a kernel interpolation is used as
an approximation of $f$. Besides it proves that the interpolation done thanks to a \textsc{maximin} design is consistent
since $\delta_{\bX_N}$ tends to $0$ for a sequence $(\bX_N)_{N\in \mathbb{N}}$ of \textsc{maximin} designs of respectively $N$ points.

\section{Computing \textsc{maximin} designs}

In this Section, we propose an algorithm to provide a \textsc{maximin} design with $N$ points in any set $E$ enclosed in a bounded set.
It is based on a simulated annealing method. It aims at finding the global minimum of the function $U:E^N\rightarrow \mathbb{R}_+$,
$U(\bX)=\mbox{diam}(E)-\delta_{\bX}$ where $\mbox{diam}(E)$ is the diameter of the set $E$ ($\mbox{diam}(E)=\max_{\bx,\bx'\in E}\Vert \bx-\bx'\Vert$).
It is obvious that to minimize $U$ is equivalent to maximize $\delta :\bX\mapsto \delta_{\bX}$. 

The initialization step consists of simulating uniformly a lot of points in the domain $E$ 
(using sampling rejection) and of calculating the corresponding empirical covariance matrix
denoted by $\Sigma$. At the end of the initialization step, we randomly keep $N$ points, denoted by $\bX^{(0)}=\{\bx_1^{(0)},\ldots,\bx_N^{(0)}\}$. 
An inverse cooling schedule $\beta:t\mapsto \beta_t$ (i.e. $(\beta_t)_t$ is an increasing positive sequence and $\lim_{t\rightarrow \infty}\beta_t=\infty$)
is chosen in order to ensure the convergence of the algorithm.
The paramater $\tau$ is a variance parameter which is allowed to change during the iterations but, at each iteration,
$\tau$ is such that $\tau_0\ge\tau\ge\tau_{min}$. 
A paramater $\gamma>0$ is fixed to be a very small integer to prevent from numerical problems.
We propose to iterate the following steps, for $t=1,\ldots$:

\vspace{1cm} \hrule
\begin{algo} 
\label{vrairecuit}
\begin{sffamily}
~
\begin{enumerate}
\item A pair of points $(\bx_i^{(t)},\bx_j^{(t)})$ is drawn in $\bX^{(t)}$ according to a multinomial distribution with probabilities proportional to $1/(\Vert \bx_i -\bx_j\Vert+\gamma)$ ;
\item One of the two points is chosen with probability $\frac{1}{2}$, it is denoted by $\bx_k^{(t)}$ ;
\item \label{proposalstep} A constraint Gaussian random walk is used to propose a new point :
$$
\bx_k^{prop}\sim \mathcal{N}^E_d(\bx_k^{(t)},\tau \Sigma)\,
$$
$\bx_k^{prop}$ is constrained to belong to $E$. The proposed design is denoted by
$\bX^{prop}=\{\bx_1^{(t)},\dots,\bx_{k-1}^{(t)},\bx_k^{prop},\bx_{k+1}^{(t)},\dots,\bx_N^{(t)}\}$;
\item \label{acceptationstep} $\bX^{(t+1)}=\bX^{prop}$ with probability
$$
\min\left(1,\exp{\left(-\beta_t(U(\bX^{prop})-U(\bX^{(t)}))\right)}\frac{q_{\tau,k}(\bX^{prop},\bX^{(t)})}{q_{\tau,k}(\bX^{(t)},\bX^{prop})}\right)\,,
$$
otherwise $\bX^{(t+1)}=\bX^{(t)}$.
\end{enumerate}
\end{sffamily}
\bigskip \hrule
\end{algo}

\vspace{0.5cm} The idea behind this proposal is to force the pairs of points which are very close to be more distant.

In order to explicit the proposal kernel $Q_{\tau}(\bX,d\bY)$ where $d\bY$ is an infinitesimal neighborhood of the state $\bY$,
let us introduce some notations:
\begin{itemize}
\item $d_{i,j}^{\bX}=1/(\Vert \bx_i -\bx_j\Vert+\gamma)$,
\item $D^{\bX}=\sum_{k,l:k< l} d_{k,l}^{\bX}$,
\item $\phi(.|\mu,S)$ denotes the Gaussian pdf with mean $\mu$ and covariance matrix $S$,
\item $G_{\mu,S}=\int_ E\phi(\by|\mu,S)d\by$ denotes the normalization constant associated to $\phi(.|\mu,S)$ on the domain $E$.
\end{itemize}
Let $\lambda$ denote the Lebesgue measure on the compact set $E$ ($\lambda(d\bx) =\I_E(\bx)\text{Leb}(d\bx)$ where
$\text{Leb}$ is the Lebesgue measure on $\mathbb{R}^d$) and let, for any $\xi$, $\text{Dir}_{\xi}(\cdot)$ denote the Dirac measure with mass at $\xi$.
For a given $\bX=\{\bx_1,\ldots,\bx_N\}\in E^N$, let $\bX_{-i}$ be such that
$\bX_{-i}=\{\bx_1,\ldots,\bx_{i-1},\bx_{i+1},\ldots,\bx_N\}$.
The proposal kernel reads as, for $\bX\in E^N,\ \bY\in (\mathbb{R}^d)^N$,
\begin{equation*}
\label{propkernelA1}
Q_{\tau}(\bX,d\bY)=\sum_{i=1}^N q_{\tau,i}(\bX,\bY)\lambda(d\by_i)\text{Dir}_{\bX_{-i}}(d\bY_{-i})\,,
\end{equation*}
where for $i=1,\ldots,N$,
$$
q_{\tau,i}(\bX,\bY)=\phi(\by_i|\bx_i,\tau\Sigma)G_{\bx_i,\tau\Sigma}^{-1}\left(\sum_{j:j\neq i} \frac{1}{2}\frac{d_{i,j}^{\bX}}{D^{\bX}}\right)\,.
$$

Let us now describe the global kernel associated to Algorithm \ref{vrairecuit}.
It obviously depends on the parameters $\beta$ and $\tau$. It reads as,
\begin{eqnarray*}
K^{A1}_{\beta,\tau}(\bX,d\bY)& = & \left\{\sum_{i=1}^N  a_{\beta,\tau,i}(\bX,\bY)q_{\tau,i}(\bX,\bY)\lambda(d\by_i)\text{Dir}_{\bX_{-i}}(d\bY_{-i})\right\} \\
                             &   & +\left(1-\int_{E^N} \sum_{i=1}^N a_{\beta,\tau,i}(\bX,\bZ)q_{\tau,i}(\bX,\bZ)\lambda(d\bz_i)\text{Dir}_{\bX_{-i}}(d\bZ_{-i})\right)\text{Dir}_{\bX}(d\bY)\,,
\end{eqnarray*}
where $a_{\beta,\tau,i}(\bX,\bY)=\min\left(1,\frac{\exp(-\beta(U(\bY)))q_{\tau,i}(\bY,\bX)}{\exp(-\beta(U(\bX)))q_{\tau,i}(\bX,\bY)}\right)$,
for $i=1,\ldots,N$.

For $\beta$ and $\tau$ fixed, the algorithm we propose is a random scan Metropolis-within-Gibbs algorithm
\citep{roberts:rosenthal:2006} for which the selection probabilities depend on the current state of the Markov chain.
Let $\lambda_N$ denote the Lebesgue measure on the compact set $E^N$ ($\lambda_N(d\bX) =\I_{E^N}(\bX)\text{Leb}_N(d\bX)$ where
$\text{Leb}_N$ is the Lebesgue measure on $(\mathbb{R}^d)^N$). The target distribution that corresponds to our random scan
Metropolis-within-Gibbs is the Gibbs measure defined by 
$$
\mu_{\beta}(d\bX)=\exp(-\beta U(\bX))Z_{\beta}^{-1}\lambda_N(d\bX)\,,
$$
where $Z_{\beta}=\int e^{-\beta U(\bY)} \lambda_N(d\bY)$.
In the presentation given by \citet{bartoli:delmoral:2001}, 
the simulated annealing algorithm is based on a Metropolis-Hastings algorithm \citep{hastings:1970} and
it is assumed that there exists a measure for which the proposal kernel is reversible. In that case,
if the target distribution is defined according this measure, the ratio of proposal densities does not appear in the
acceptance rate. In our case, there does not exist a reversible measure for $Q_\tau$, that is why the ratio of the proposal densities is needed. 

We will show the convergence of Algorithm \ref{vrairecuit} following the proof given in \cite{bartoli:delmoral:2001}.
Some Markov chains convergence results are used. Indeed, at a fixed temperature, the Markov chain tends to a stationnary
distribution which is the Gibbs measure. As the temperature decreases, the Gibbs measure concentrates on the global extremum of the
function. There are typically two properties that are required in the proof: the irreducibility and the invariance of the transition
kernel with respect to the Gibbs measure.
As already explained, for $\beta$ and $\tau$ fixed, this scheme is a random scan Metropolis-within-Gibbs algorithm with
$\mu_\beta$ as target. Indeed, using the detailed balance condition, we can easily verified that $\mu_\beta$ is
$K^{A1}_{\beta,\tau}$-invariant. The irreducibility of this scheme comes from the following result.
\begin{proposition}
\label{irreducibility}
For all non-decreasing sequence $0\le\beta_1\le\ldots\le\beta_N$ and for a
sequence $(\tau_i)_{1\le i\le N}$ such $\tau_0\ge\tau_i\ge\tau_{min}$, we have $\forall (\bX,A)\in E^N\times \mathbb{B}(E^N)$
$$
(K^{A1}_{\beta_1,\tau_1}\cdots K^{A1}_{\beta_N,\tau_N})(\bX,A)\ge \alpha_{A1} e^{-N\beta_N osc(U)}\lambda_N(A)
$$
where $\alpha_{A1}>0$ and $osc(U)$ is the smallest positive number $h$ such that for all $\bX$, $\bY$ in $E^N$, $U(\bY)-U(\bX)\le h$.
\end{proposition}

\Preuve \\
Let us prove first that, for all $\bX\in E^N$ and $i=1,\ldots,N$, $q_{\tau,i}(\bX,.)\ge q_{min}>0$ and
 $q_{\tau,i}(\bX,.)\le q_{max}$, $Q_{\tau}(\bX,.)$-almost everywhere on $E^N$. \\
The fact that $q_{\tau}(\bX,.)\le q_{max}$ is true since the normalization constants are lower-bounded, the Gaussian densities are uniformly
bounded since $\tau_0\ge \tau \ge \tau_{min}>0$ and all the other terms can be upper bounded by $1$.\\
The other assertion is only true $Q_{\tau}(\bX,.)$-almost everywhere on $E^N$. It means that the lower bound on $q_{\tau,i}(\bX,\bY)$
holds  when $\bX$ and $\bY$ are such that $\bX_{-i}=\bY_{-i}$ and are both in $E^N$. \\
The following lower bounds are used:
\begin{itemize}
\item $G_{\bx_i,\tau\Sigma}^{-1}\ge1$,
\item $\sum_{j:j\neq i} \frac{1}{2}\frac{d_{i,j}^{\bX}}{D^{\bX}}\ge \frac{\gamma}{N(\mbox{diam}(E)+\gamma)}$,
\item $\phi(\by|\bx,\tau\Sigma)\ge\frac{1}{(2\pi)^{d/2}|\Sigma|^{1/2}\tau_0^{d/2}}\exp\left(-\frac{1}{2}\tau_{min}^{-1}\mbox{diam}(E)^2\xi\right)$
where $\xi$ is the largest eigenvalue of $\Sigma^{-1}$.
\end{itemize}
$q_{min}>0$ is found by multiplying these expressions and, for any $i=1,\ldots,N$,
it is a lower bound of $q_{\tau,i}(\bX,\bY)$ which does not depend on $\tau$ and on the states
if $\bX\in E^N$ and $\bY\in E^N$ have at least $N-1$ points in common. 

\noindent By definition of $osc(U)$, for all $\bX\in E^N$ and for all $Q_{\tau}(\bX,.)$-almost everywhere $\bY\in E^N$,
$$
a_{\beta,\tau,i}(\bX,\bY) \ge e^{-\beta osc(U)}\frac{q_{min}}{q_{max}}\,.
$$
Thus for $(\bX,A)\in E^N\times \mathbb{B}(E^N)$, $K^{A1}_{\beta,\tau}(\bX,A)\ge e^{-\beta osc(U)}\frac{q_{min}}{q_{max}}Q_{\tau}(\bX,A)$.

\noindent Then, for all non-decreasing sequence $0\le\beta_1\le\ldots\le\beta_N$ and for a
sequence $(\tau_i)_{1\le i\le N}$ such $\tau_0\ge\tau_i\ge\tau_{min}$, we have $\forall (\bX,A)\in E^N\times \mathbb{B}(E^N)$ 
\begin{eqnarray*}
(K^{A1}_{\beta_1,\tau_1}\cdots K^{A1}_{\beta_N,\tau_N})(\bX,A)&\ge& e^{-(\beta_1+\cdots+\beta_N)osc(U)}
\left(\frac{q_{min}}{q_{max}}\right)^N(Q_{\tau_1}\cdots Q_{\tau_N})(\bX,A) \\
&\ge & e^{-N\beta_N osc(U)}\left(\frac{q_{min}}{q_{max}}\right)^N(Q_{\tau_1}\cdots Q_{\tau_N})(\bX,A) \\
&\ge & e^{-N\beta_N osc(U)}\left(\frac{q_{min}}{q_{max}}\right)^N q_{min}^N \lambda_N(A)\,.
\end{eqnarray*}
\Fin

\noindent $U:E^N\rightarrow \mathbb{R}_+$ is lower bounded with respect to $\lambda_N$ (the Lebesgue measure on the compact set $E^N$). We use the following notation
$m=\sup_a[a;\lambda_N(\{\bX;U(\bX)<a\})=0]$, by definition $\lambda_N(\{\bX;U(\bX)<m\})=0$. \\
Moreover, for all $\epsilon>0$, we define $U_{\lambda_N}^{\epsilon}=\{\bX\in E^N;U(\bX)\le m+\epsilon \}$ which is clearly such that
$\lambda_N(U_{\lambda_N}^{\epsilon})>0$ and $U_{\lambda_N}^{\epsilon,c}=\{\bX;\ U(\bX)>m+\epsilon \}$. 

\noindent As stated in \cite{bartoli:delmoral:2001}
$$\forall \epsilon>0,\ \ \ \lim_{\beta\rightarrow \infty}\mu_{\beta}(U_{\lambda_N}^{\epsilon})=1\,. $$
Following the proof of Theorem 4.3.16 in \cite{bartoli:delmoral:2001} and
using Proposition \ref{irreducibility}, we obtain the convergence of this algorithm.

\begin{theoremeeng}
\label{cvrecuit}
If the sequence $(\tau_n)_{n\ge 0}$ is such that $\forall n\ge 0,\ \tau_0\ge\tau_n\ge\tau_{min}>0$ and if 
$$
\beta_n=\frac{1}{C}\log(n+e),\ \ \ C>N osc(U)\,,
$$
we get
$$
\forall \epsilon>0, \ \ \ \lim_{n\rightarrow \infty}\mathbb{P}_{\eta}(\bX^{(n)}\in U_{\lambda_N}^{\epsilon})=1
$$
where $\{\bX^{(n)};n\ge 0\}$ denotes
the random sequence we get from Algorithm \ref{vrairecuit} with an initial probability distribution
$\eta$ on $E^N$.
\end{theoremeeng}
Unfortunately, the function $U$ is not regular enough to estimate the convergence speed. 

\vspace{-0.5cm} \section{Variants of the algorithm}
In the case where $E$ is not explicit, the normalization constant $G_{m,S}$ of a Gaussian distribution with mean $m$ and covariance matrix $S$
cannot be computed. Hence, the ratio of densities of proposal kernels is not tractable.  
In that case, we first propose to use as a proposal an unconstrained Gaussian random walk.
The steps \ref{proposalstep} and \ref{acceptationstep} of Algorithm \ref{vrairecuit} are modified.

\vspace{0.5cm} \hrule
\begin{algo}
\label{recuithorsdomaine} 
The first steps until step \ref{proposalstep} are the same.\\
Step \ref{proposalstep} is replaced with
\begin{sffamily}
\begin{enumerate}
\item[\textit{3bis.}] A Gaussian random walk is used to propose a new point : 
$$
\bx_k^{prop}\sim \mathcal{N}_d(\bx_k^{(t)},\tau \Sigma)\,.
$$
\end{enumerate}
\end{sffamily}
And step \ref{acceptationstep} is replaced with
\begin{sffamily}
\begin{enumerate}
\item[\textit{4bis.}] If $\bX^{prop}\in E^N$, $\bX^{(t+1)}=\bX^{prop}$ with probability
$$
\min\left(1,\exp{\left(-\beta_t(U(\bX^{prop})-U(\bX^{(t)})))\right)}\frac{\tilde q_{\tau,k}(\bX^{prop},\bX^{(t)})}{\tilde q_{\tau,k}(\bX^{(t)},\bX^{prop})}\right)\,,
$$
otherwise $\bX^{(t+1)}=\bX^{(t)}$.
\end{enumerate}
\end{sffamily}
\bigskip \hrule
\end{algo}

The proposal kernel corresponding to this algorithm where the Gaussian random walk is not constraint to remain in the domain $E$ reads as: 
for any $\bX\in E^N,\ \bY \in (\mathbb{R}^d)^N$,
$$
Q_{\tau}(\bX,d\bY)=\sum_{i=1}^N \tilde q_{\tau,i}(\bX,\bY)\lambda(d\by_i)\text{Dir}_{\bX_{-i}}(d\bY_{-i})\,,
$$
where for $i=1,\ldots,N$,
$$
\tilde q_{\tau,i}(\bX,\bY)=\phi(\by_i|\bx_i,\tau\Sigma ) )\left(\sum_{j:j\neq i} \frac{1}{2}\frac{d_{i,j}^{\bX}}{D^{\bX}}\right)\,.
$$
As for Algorithm \ref{vrairecuit}, if $(\tau_n)_{n\ge 0}$ is such that $\forall n\ge 0,\ \tau_0\ge\tau_n\ge\tau_{min}>0$ and if 
$\beta_n=\frac{1}{C}\log(n+e)$ with $C>N osc(U)$, the random sequence we get from Algorithm 2 gives
$\forall \epsilon>0, \ \ \ \lim_{n\rightarrow \infty}\mathbb{P}_{\eta}(\bX^{(n)}\in U_{\lambda_N}^{\epsilon})=1$.
However, since a point can be proposed outside of the domain $E$, this algorithm can suffer from a lack of efficiency.  
Another solution is to use the first algorithm without the ratio of densities of proposal kernels. 

\newpage

\hrule
\begin{algo}
\label{recuitechange} 
The first steps until step \ref{acceptationstep} are the same than in Algorithm \ref{vrairecuit}.\\
Step \ref{acceptationstep} is replaced with
\begin{sffamily}
\begin{enumerate}
 \item[\textit{4ter.}] 
 $\bX^{(t+1)}=\bX^{prop}$ with probability
$$
\min\left(1,\exp{\left(-\beta_t(U(\bX^{prop})-U(\bX^{(t)})))\right)}\right)\,,
$$
otherwise $\bX^{(t+1)}=\bX^{(t)}$.
\end{enumerate}
\end{sffamily}
\bigskip \hrule
\end{algo}
The global kernel associated to Algorithm \ref{recuitechange} is
$$
K^{A3}_{\beta,\tau}(\bX,d\bY)=b_{\beta,\tau}(\bX,\bY)Q_{\tau}(\bX,d\bY)+\left(1-\int_{E^N} b_{\beta,\tau}(\bX,\bZ)Q_{\tau}(\bX,d\bZ)\right)\delta_{\bX}(d\bY)\,,
$$
where $b_{\beta,\tau}(\bX,\bY)=\min\left(1,\frac{\exp(-\beta U(\bY))}{\exp(-\beta U(\bX))}\right)$.
The measure $\mu_\beta$ is not $K^{A3}_{\beta,\tau}$-invariant. Hence, we cannot use the Markov chain convergence theory
to obtain a result similar to Theorem \ref{cvrecuit}. However, $\mu_\beta$ is $K^{A3}_{\beta,\tau}$-irreducible and
we can easily state the following proposition.
\begin{proposition}
\label{irreducibility3}
For all non-decreasing sequence $0\le\beta_1\le\ldots\le\beta_N$ and for a
sequence $(\tau_i)_{1\le i\le N}$ such $\tau_0\ge\tau_i\ge\tau_{min}$, we have $\forall (\bX,A)\in E^N\times \mathbb{B}(E^N)$
$$
(K^{A3}_{\beta_1,\tau_1}\cdots K^{A3}_{\beta_N,\tau_N})(\bX,A)\ge \alpha_{A3} e^{-N\beta_N osc(U)}\lambda_N(A)
$$
where $\alpha_{A3}>0$.
\end{proposition}
As shown in \cite{locatelli:1996}, this proposition leads to the fact that a design reaching a neighborhood
$U_{\lambda_N}^{\epsilon}$ of a global maximum of $\delta_\bX$ can be achieved in a finite number of iterations almost surely
using Algorithm \ref{recuitechange}.
\begin{proposition}
For any $\epsilon>0$, if, $\forall n\in\mathbb{N}$, $\beta_n\le\frac{1}{C}\log(n+e)$ with $C>N osc(U)$
the expected time until the first visit in $U_{\lambda_N}^{\epsilon}$ is finite. 
\end{proposition}
\Preuve \\
The expected time until the first visit in $U_{\lambda_N}^{\epsilon}$ is equal to 
$$
\sum_{k=1}^{\infty}k\mathbb{P}(\bX^{(1)},\ldots ,\bX^{(k)}\notin U_{\lambda_N}^{\epsilon}|\bX^{(0)}\notin
U_{\lambda_N}^{\epsilon})\times \mathbb{P}(\bX^{(k+1)}\in U_{\lambda_N}^{\epsilon}|\bX^{(0)},\ldots ,\bX^{(k)}
\notin U_{\lambda_N}^{\epsilon})\,.
$$
The aim is to find an upper bound in order to show that it is finite. The second probability in the argument of
the series is limited from above with one. The first probability in the argument of the series is the probability
of never visiting $U_{\lambda_N}^{\epsilon}$ in the first $k$ steps. \\
It can also be written as:
$$
\mathbb{P}(\bX^{(1)},\ldots ,\bX^{(N)}\notin U_{\lambda_N}^{\epsilon}|\bX^{(0)}\notin U_{\lambda_N}^{\epsilon})\times
\cdots\times\mathbb{P}(\bX^{(\lfloor k/N\rfloor N)},\ldots,\bX^{(k)}\notin U_{\lambda_N}^{\epsilon}|
\bX^{(\lfloor k/N\rfloor N-1)},\ldots,\bX^{(0)}\notin U_{\lambda_N}^{\epsilon})\,.
$$ 
Thanks to Proposition \ref{irreducibility3}, it holds that
$$
\mathbb{P}(\text{at least one visit in } 
U_{\lambda_N}^{\epsilon} \text{ in the first } N \text{steps})\ge 
\mathbb{P}(\bX^{(N)}\in U_{\lambda_N}^{\epsilon})\ge\alpha_{A3}\lambda_N(U_{\lambda_N}^{\epsilon})\exp(-\beta_N N osc(U))\,.
$$
Thus,
$$\mathbb{P}(\bX^{(1)},\ldots,\bX^{(N)}\notin U_{\lambda_N}^{\epsilon}|\bX^{(0)}\notin U_{\lambda_N}^{\epsilon})\le 1-\alpha_{A3}\lambda_N(U_{\lambda_N}^{\epsilon})\exp(-\beta_N N osc(U))\,.$$
And in a similar way,
$$\mathbb{P}(\bX^{(iN+1)},\ldots,\bX^{((i+1)N)}\notin U_{\lambda_N}^{\epsilon}|\bX^{(0)},\ldots,\bX^{(iN)}\notin U_{\lambda_N}^{\epsilon})\le 1-\alpha_{A3}
\lambda_N(U_{\lambda_N}^{\epsilon})\exp(-\beta_{N(i+1)} N osc(U))\,.$$
Hence, the expected time before the first visit in $U_{\lambda_N}^{\epsilon}$ can be bounded from above by 
$$
\sum_{k=1}^{\infty}k\prod_{i=1}^{\lfloor k/N\rfloor}\left( 1-\alpha_{A3}\lambda_N(U_{\lambda_N}^{\epsilon})\exp(-\beta_{N(i+1)} N osc(U))\right)\,.
$$

As $\log(1-2x)<-x$ if $0<x<1/2$, the previous sum is bounded by
$$
\sum_{k=1}^{\infty}k\exp\left(-\sum_{i=1}^{\lfloor k/N\rfloor}\frac{\alpha_{A3}\lambda_N(U_{\lambda_N}^{\epsilon})}{2}\exp(-\beta_{N(i+1)} N osc(U))\right)\,.
$$
If $\beta_k$ is chosen such that $\beta_n=\frac{1}{C}\log(n+e),\ \ \ (C>N osc(U))$, the sum becomes
$$\sum_{k=1}^{\infty}k\exp\left(-\frac{\alpha_{A3}\lambda_N(U_{\lambda_N}^{\epsilon})}{2}\sum_{i=1}^{\lfloor k/N\rfloor}\left(\frac{1}{(i+1)N}\right)^{Nosc(u)/C}\right)\,,$$
which can be bounded above by
$$\sum_{k=1}^{\infty}k\exp\left(-\frac{\alpha_{A3}\lambda_N(U_{\lambda_N}^{\epsilon})}{2}\lfloor k/N\rfloor\left(\frac{1}{(k+N)}\right)^{Nosc(u)/C}\right)\,,$$
which is a convergent series.
\Fin
Since the best design ever found during the iterations is saved, the previous proposition provides a theoretical guarantee
for Algorithm \ref{recuitechange}. Moreover, as a direct consequence, the Markov chain defined by $\bY^{(n)}=\bX^{(t_n)}$ with $t_n=\arg\min_{1\le t\le n} U(\bX^{(t)})$
is such that $\lim_{n\rightarrow \infty}\mathbb{P}_{\eta}(\bY^{(n)}\in U_{\lambda_N}^{\epsilon})=1$.
In practice, $n$ is finite and the chosen design is $\bY^{(n)}$ and not $\bX^{(n)}$. Also, we can consider
that the previous convergence result is sufficient. However, this kind of result can be obtained with any algorithm producing a Markov chain
which well visits the space of states even if the temperature is fixed! Algorithm \ref{recuitechange} directly derives from Algorithm \ref{vrairecuit} and we can expect
that they have similar behaviors.

\section{Numerical illustrations}

First, the three algorithms are tested on three different toy cases: a design with $100$ points in $[0,1]^2$, a design with $250$ points in $[0,1]^5$
and a design with $400$ points in $[0,1]^8$. In these hypercubic cases, the normalization constants can be computed and Algorithm \ref{vrairecuit}
can be used. In each case, $100$ calls are made to one million iterations of each algorithm.
We observed that the chains produced by the algorithms remain quite stationnary 
after one million iterations. The inverse cooling schedule is $\beta_n=(1/T_0)\log(n)$ and the variance schedule is $\tau_n=\tau_0/\sqrt{n}$.

In order to choose $T_0$, a lot of designs with $N$ points can be drawn uniformly in $E$. Then, a median of $\delta_\bX$, the minimum distance
between pair of points in these designs, is computed. Thus, it is a mean to access to an order of magnitude of $\delta_\bX$ when $\bX$ is uniformly distributed.
A fraction of this value is a good choice for $T_0$ according to our tries. Note that it is much lower than the one required in the convergence theorem.

For $\tau_0$, we suggest to use $\tau_0=\mbox{Vol}(E)/N^{1/d}$ where $\mbox{Vol}(E)$ is the volume of $E$ or an upper bound of this volume.
Clearly, $\tau_0$ which parametrizes the random walk variance should not exceed $\mbox{Vol}(E)$ and the previous formula is derived from analogy with a grid.
For a $d$-dimensional space, the number of points in a grid reads as $N=k^d$ where $k$
is an integer which corresponds to the number of projected points on each axis. Thus, it seems reasonable to divide the volume of
the domain by $k$ or more generally by $N^{1/d}$ to ensure a good exploration of the space.

\begin{figure}[ht]
\centering
\includegraphics[width=0.8\textwidth]{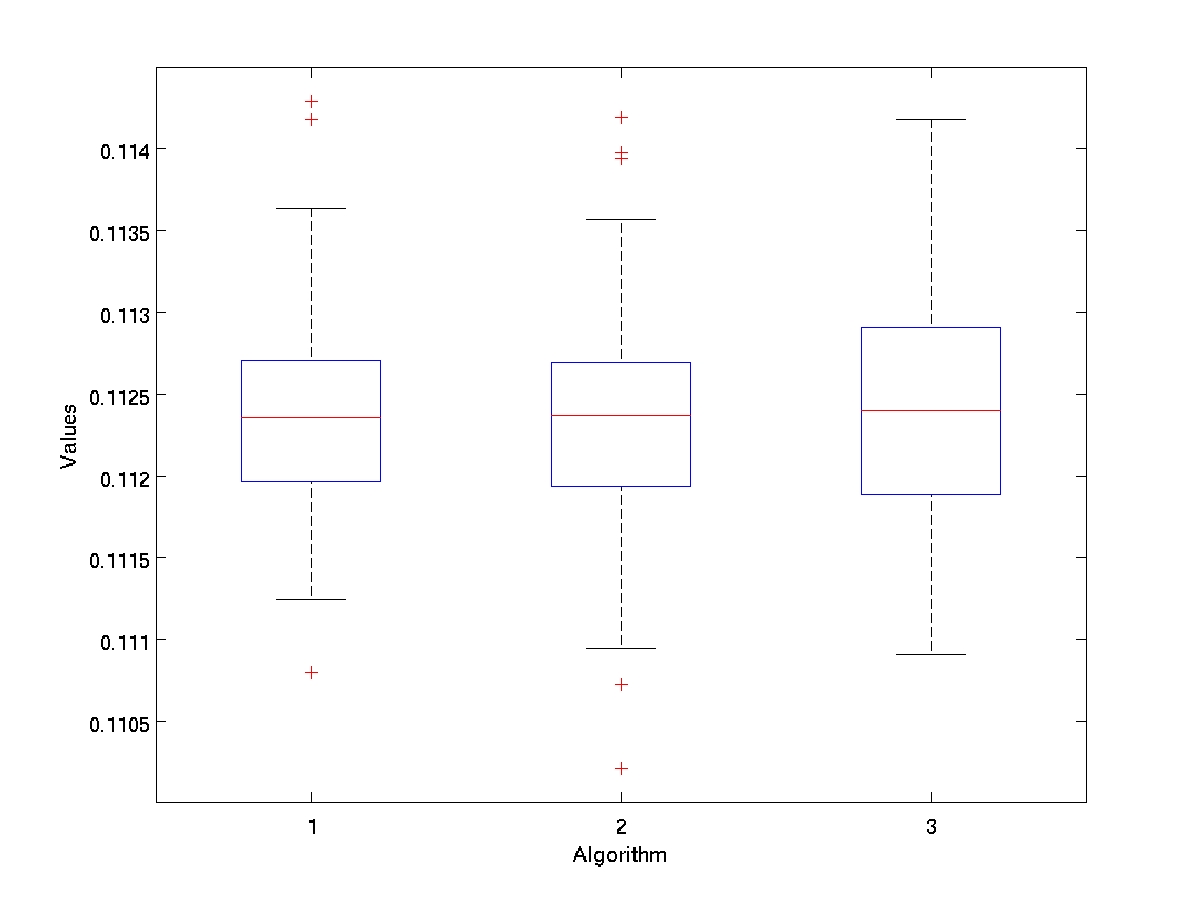}
\caption{\label{fig1} Case of a design of $100$ points in $[0,1]^2$}
\end{figure}

\begin{figure}[ht]
\centering
\includegraphics[width=0.8\textwidth]{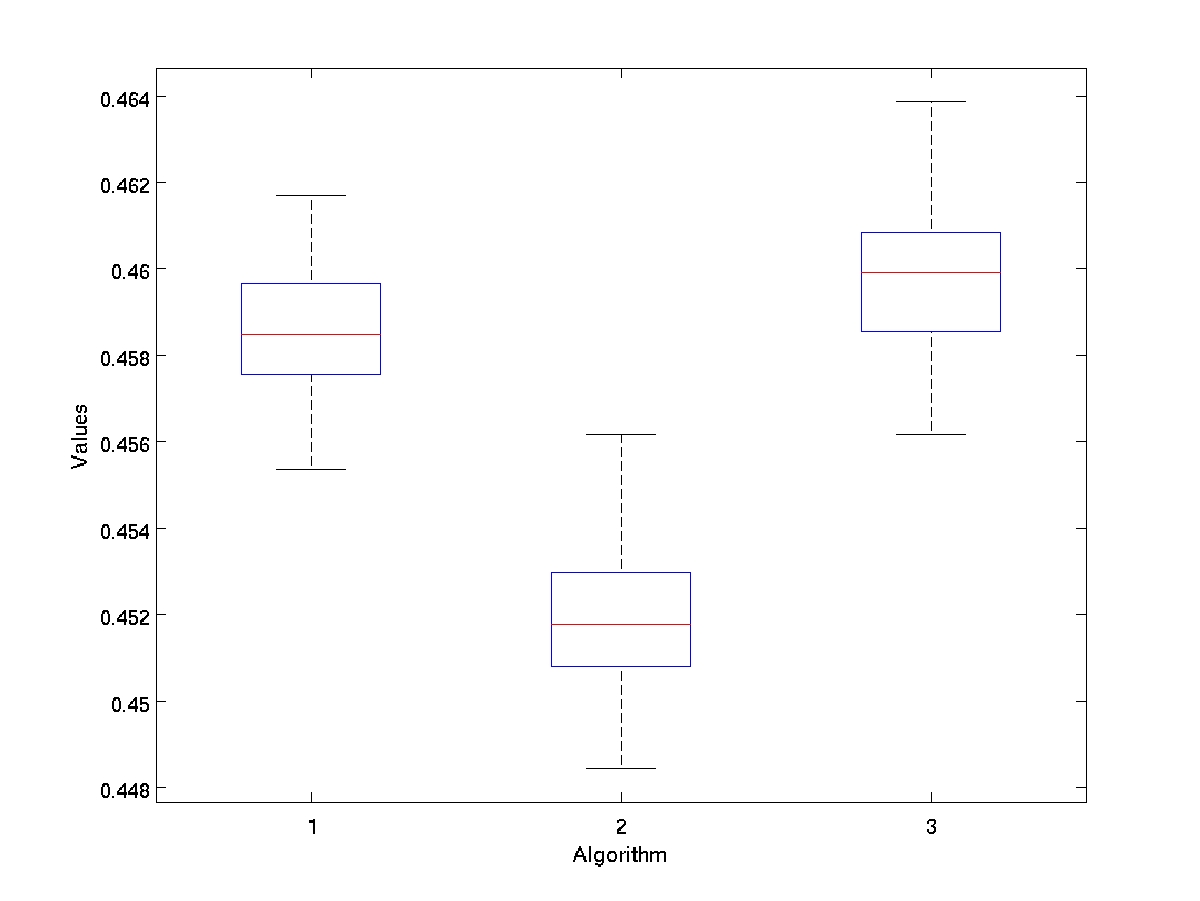}
\caption{\label{fig2} Case of a design of $250$ points in $[0,1]^5$}
\end{figure}

\begin{figure}[ht]
\centering
\includegraphics[width=0.8\textwidth]{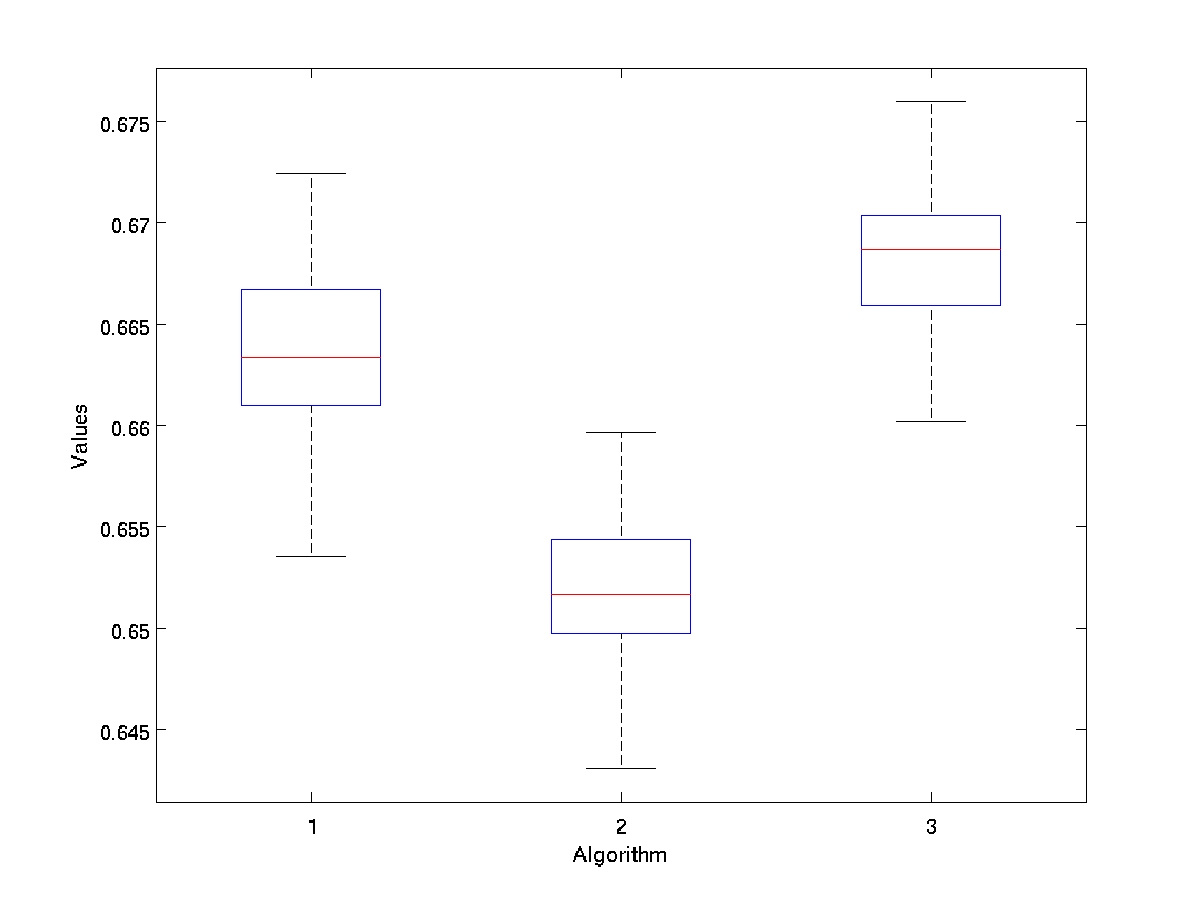}
\caption{\label{fig3} Case of a design of $400$ points in $[0,1]^8$}
\end{figure}

Figures \ref{fig1}, \ref{fig2} and \ref{fig3} present the results.
For each algorithm, the boxplots of the best solutions to the maximization of $\delta_\bX$ over one million iterations (boxplots are constructed using
100 replicates) are given. Algorithms \ref{vrairecuit} and \ref{recuitechange} give the best results. Algorithm \ref{recuithorsdomaine} suffers from the fact that the proposal
can be outside of the domain. The computation of $U$ is the most time consuming step of these algorithms, that is why the comparison is based
on the number of iterations.

Other cooling schedules than the ones which have theoretical guarantees can be tried. It seems that they can lead to 
satisfying results which are even better than the ones obtained with the
$\log$ schedule. Since the results depend too much on the examples, it is quite hard to state a general rule. However,
a schedule $\beta_n=1/T_0\sqrt{n}$ is robust to a bad choice in $T_0$ and a schedule $\tau_n=\tau_0/\sqrt{n}$ performs quite well.
The variance decreasing schedule is set to freeze for a given $n$, thus it satisfies the boundedness assumption of the theoretical
results.

Finally, in the domain $E_T=\{(x_1,x_2)\in [0,1]^2: x_1>x_2\}$, four strategies are compared to obtain
a design with $100$ points: taking a design whose points are realizations of a uniform distribution on 
$E_T$, taking a \textsc{lhs-maximin} design in  $[0,1]^2$ with $200$ points thanks to the algorithm of
\citet{morris:mitchell:1995} and keeping the subset which is in $E_T$ only (it gives what we call a truncated \textsc{lhs-maximin} design),
using a Sobol' sequence of $100$ points constrained to be in $E_T$, making use of Algorithm \ref{recuitechange}.
Table \ref{maximintriangle} displays some statistics on values of $\delta_{\bX}$ for $100$ replicates.
Only the mean is given for the Sobol' sequence since it is a deterministic strategy. The 
truncated \textsc{lhs-maximin} strategy provides designs with approximately $100$ points (between $93$ and $107$
on the $100$ replicates).
\begin{table}[h]
\begin{center}
 \begin{tabular}{|c|cccc|}
\hline
       & Mean & Variance & Min & Max \\
\hline
Uniform&  0.0048 & $8.2\cdot 10^{-6}$     &$4.0\cdot 10^{-4} $& 0.013\\
\hline
Truncated \textsc{lhs-maximin} & 0.034& $8.2\cdot 10^{-6}$ & 0.025&0.039 \\
\hline
Sobol' sequence &0.011& N/A&N/A &N/A\\
\hline
Algorithm \ref{recuitechange}& 0.080  &$7.8\cdot 10^{-8}$  & 0.079   &0.081\\
\hline
\end{tabular}
\caption{Comparison of designs in $E_T$ based on the values of $\delta_{\bX}$ for $100$ replicates}
\label{maximintriangle}
\end{center}
\end{table}

\section{Application to a simulator of an aircraft engine}

The behavior of an aircraft engine is described by a numerical code. A run of the code determines if the given flight conditions are acceptable 
and, provided they are, computes the corresponding outputs.
The function which associates the outputs to the flight conditions is denoted by $f$.
It is accessible only through runs of the code. It is a black box function and a run is time-consuming.
A thousand calls to the code are run in ten minutes. The goal is to incorporate the modelization of the engine in a global model of an aircraft for
a preliminary design study. Since the simulator of the engine is too burdensome,
we are asked to compute an approximation of $f$ which can be included in the global model.

The acceptable flight conditions represent the domain of definition of $f$, denoted by $E$.
Outside $E$, the code cannot provide outputs since the conditions are physically impossible or the
code encounters convergence failures. $E$ is not explicit, as explained above we have to run the code
to know if the flight conditions are acceptable. Therefore, we need to estimate $E$
(the indicator function associated to $E$). This is not our goal here.
$E$ is included in a known hypercube (lower and upper bounds are available on each of these variables). 
Using other prior information and some calls to $f$,
a binary classification tree has been built to determine an estimate of the indicator function of $E$ 
\citep{auffray:barbillon:marin:2011}.
This method works quite well and leads to a misclassification error rate around $0.5\%$. The resulting
domain is not an hypercube.

In the following case study, only the flow rate output is focused on. 
The flight conditions are described by ten variables such as altitude, speed, temperature, humidity...
A variable selection procedure has shown that only $d=8$ input variables are useful for prediction
\citep{auffray:barbillon:marin:2011}.
Hence, the considered function to be approximated is $f:E\subset \mathbb{R}^d\rightarrow \mathbb{R}$.

A {\sc \textsc{maximin}} design is drawn thanks to $10^7$ iterations of Algorithm \ref{recuitechange}.
The initial temperature $T_0$ and the initial variance $\tau_0$ were chosen as described in the previous section.
The inverse cooling shedule was $\beta_n=(1/T_0)\sqrt{n}$ and the variance schedule was constant during the first
quarter of iterations and then $\tau_n=\tau_0/\sqrt{n-10^7/4}$.

Approximations of the function $f$ are made by kernel interpolations on four different designs:
the {\sc \textsc{maximin}} design that was computed, a design whose points follow a uniform distribution on $E$,
a design obtained by truncating a Latin hypercube design of $5,000$ points defined on the hypercube domain containing $E$
and a design given by a low-discrepancy sequence (Sobol) constrained to be in $E$ \citep[see][]{bratley:fox:1988}.
The \textsc{lhs} is truncated by keeping only the points which belongs to $E$. 
The kernel interpolations are computed by the Matlab toolbox DACE \citep{lophaven:etal:2002}.
The regression functions are chosen as the polynomials with degree smaller than or equal to two and the
kernel is a generalized exponential kernel:
$$
K(\bx,\bx')=\exp\left(-\sum_{j=1}^d \theta_j|x^{(j)}-x'^{(j)}|^\nu\right)\,,
$$
where $x^{(j)},x'^{(j)}, j=1,\ldots,d$ are respectively the $j^{\text{th}}$ coordinates of $\bx, \bx'$
and $\theta_1,\ldots,\theta_d,\nu$ are parameters which are estimated using the usual maximum likelihood estimators. 
Others methods such as cross validation could be used to choose these parameters.
The results given in Section 2 only applies when the kernel is isotropic and Gaussian which means 
$\theta_1=\ldots=\theta_d=\theta$ and $\nu=2$.

The four designs are sets of approximately $1,300$ points which are included in the domain $E$
according to the estimated indicator function. For the \textsc{lhs}, we need around $5,000$ points to get approximately
$1,300$ points in $E$.

The function $f$ is computed at the points of the designs. Some points have to be removed from the designs since the code indicates
that they are not in $E$ (recall that the designs were built thanks to an estimate of $E$).  

Table \ref{comparisondesigns} provides the performances of kernel interpolations according to the designs.
The performances are evaluated on another set of $T=1,300$ points uniformly distributed in $E$ (obtained using sampling rejection)
on which the function $f$ is also computed. If $\hat f$ denotes a kernel interpolator and $\{\bz_1,\ldots,\bz_{T}\}$
is the set of test points, those quantities are reported:
\begin{itemize}
 \item the Mean Relative Error (MRE), 
$$\frac{1}{T}\sum_{i=1}^{T}\left|\frac{f(\bz_i)-\hat f(\bz_i)}{f(\bz_i)}\right| \,,$$
\item the Maximum Relative Error (MaxRE),
$$\max_{i=1,\ldots,T}\left|\frac{f(\bz_i)-\hat f(\bz_i)}{f(\bz_i)}\right| \,,$$
\item the Mean Squared Error (MSE), 
$$\frac{1}{T}\sum_{i=1}^{T}\left(f(\bz_i)-\hat f(\bz_i)\right)^2 \,.$$
\end{itemize}
Table \ref{comparisondesigns} also contains the number of points which are actually in $E$ and the minimal distance
$\delta_\bX$ between the pairs of points of the designs. To compute these distances, the designs were translated into
the hypercube $[0,1]^8$.
\begin{table}[h]
\begin{center}
 \begin{tabular}{|c|ccccc|}
\hline
       & mRE& MaxRE& MSE&Nb of Points& $\delta_{\bX}$ \\
\hline
Uniform&  0.49\% & 5.2\%     & 0.63  & 1284& 0.15\\
\hline
\textsc{lhs}    & 0.48\%  & 6.9\%     &0.73   & 1275&0.14\\
\hline
{\sc \textsc{maximin}}& 0.47\%  &   3.5\%   & 0.56   &1249& 0.33\\
\hline
Sobol' sequence &0.46\%  &  7.7\% & 0.62 &1277& 0.15\\
\hline
\end{tabular}
\caption{Comparison of performances of kernel interpolation on the different designs}
\label{comparisondesigns}
\end{center}
\end{table}

The \textsc{maximin} design makes the kernel interpolation more efficient especially
according to the MaxRE criterion (although it contains less admissible points than other designs).
As it was shown, the kernel interpolation accuracy depends sharply on the spreading
out of the points of the design. Thus, the \textsc{maximin} design which ensures that any point of $E$ is not far
from the points of the design leads to the best performances.

\section*{Acknowledgements}

The authors are grateful to Pierre Del Moral for very helpful discussions on the convergence properties of the algorithms.
This work has been supported by the Agence Nationale de la Recherche (ANR, 212, rue de Bercy 75012 Paris)
through the 2009-2012 project Big'MC.

\bibliographystyle{apalike}
\bibliography{ABM11-SA}

\begin{thebibliography}{}

\bibitem[Auffray et~al., 2011]{auffray:barbillon:marin:2011}
Auffray, Y., Barbillon, P., and Marin, J.-M. (2011).
\newblock Mod\`eles r\'eduits \`a partir d'exp\'eriences num\'eriques.
\newblock {\em Journal de la Soci\'et\'e Fran\c{c}aise de Statistique},
  152(1):89--102.

\bibitem[Bartoli and Del~Moral, 2001]{bartoli:delmoral:2001}
Bartoli, N. and Del~Moral, P. (2001).
\newblock {\em Simulation \& algorithmes stochastiques}.
\newblock C\'epadu\`es.

\bibitem[Bratley and Fox, 1988]{bratley:fox:1988}
Bratley, P. and Fox, B.~L. (1988).
\newblock Algorithm 659: {I}mplementing {S}obol's quasirandom sequence
  generator.
\newblock {\em ACM Transactions on Mathematical Software}, 14(1):88--100.

\bibitem[Bursztyn and Steinberg, 2006]{bursztyn:steinberg:2006}
Bursztyn, D. and Steinberg, D.~M. (2006).
\newblock Comparison of designs for computer experiments.
\newblock {\em Journal of Statistical Planning and Inference},
  136(3):1103--1119.

\bibitem[Chib and Greenberg, 1995]{chib:1995}
Chib, S. and Greenberg, E. (1995).
\newblock Understanding the {M}etropolis-{H}astings algorithm.
\newblock {\em The American Statistician}, 49(4):327--335.

\bibitem[Cressie, 1993]{cressie:1993}
Cressie, N. A.~C. (1993).
\newblock {\em Statistics for spatial data}.
\newblock Wiley Series in Probability and Mathematical Statistics: Applied
  Probability and Statistics. John Wiley \& Sons Inc., New York.

\bibitem[den Hertog et~al., 2006]{denhertog:kleijnen:siem:2006}
den Hertog, D., Kleijnen, J., and Siem, A. (2006).
\newblock The correct {K}riging variance estimated by bootstrapping.
\newblock {\em The Journal of the Operational Research Society},
  57(4):400--409.

\bibitem[Fang et~al., 2005]{fang:li:sudjianto:2005}
Fang, K.-T., Li, R., and Sudjianto, A. (2005).
\newblock {\em Design and Modeling for Computer Experiments (Computer Science
  \& Data Analysis)}.
\newblock Chapman \& Hall/CRC.

\bibitem[Hastings, 1970]{hastings:1970}
Hastings, W. (1970).
\newblock Monte {C}arlo {S}ampling {M}ethods {U}sing {M}arkov {C}hains and
  {T}heir {A}pplications.
\newblock {\em Biometrika}, 57(1):97--109.

\bibitem[Johnson et~al., 1990]{johnson:etal:1990}
Johnson, M.~E., Moore, L.~M., and Ylvisaker, D. (1990).
\newblock Minimax and maximin distance designs.
\newblock {\em Journal of Statistical Planning and Inference}, 26(2):131--148.

\bibitem[Joseph, 2006]{joseph:2006}
Joseph, V.~R. (2006).
\newblock Limit kriging.
\newblock {\em Technometrics}, 48(4):458--466.

\bibitem[Koehler and Owen, 1996]{koehler:owen:1996}
Koehler, J.~R. and Owen, A.~B. (1996).
\newblock Computer experiments.
\newblock In {\em Design and analysis of experiments}, volume~13 of {\em
  Handbook of Statistics}, pages 261--308. North-Holland, Amsterdam.

\bibitem[Laslett, 1994]{laslett:1994}
Laslett, G.~M. (1994).
\newblock Kriging and splines: an empirical comparison of their predictive
  performance in some applications.
\newblock {\em Journal of the American Statistical Association},
  89(426):391--409.

\bibitem[Li and Sudjianto, 2005]{li:sudjianto:2005}
Li, R. and Sudjianto, A. (2005).
\newblock Analysis of computer experiments using penalized likelihood in
  {G}aussian {K}riging models.
\newblock {\em Technometrics}, 47(2):111--120.

\bibitem[Locatelli, 1996]{locatelli:1996}
Locatelli, M. (1996).
\newblock Convergence {P}roperties of {S}imulated {A}nnealing for {C}ontinuous
  {G}lobal {O}ptimization.
\newblock {\em Journal of Applied Probability}, 33(4):1127--1140.

\bibitem[Lophaven and Sondergaard, 2002]{lophaven:etal:2002}
Lophaven, N.S., N.~H. and Sondergaard, J. (2002).
\newblock {DACE}, a {M}atlab {K}riging toolbox.
\newblock Technical Report IMM-TR-2002-12, DTU.

\bibitem[Madych and Nelson, 1992]{madych:nelson:1992}
Madych, W.~R. and Nelson, S.~A. (1992).
\newblock Bounds on multivariate polynomials and exponential error estimates
  for multiquadric interpolation.
\newblock {\em Journal of Approximation Theory}, 70(1):94--114.

\bibitem[Matheron, 1963]{matheron:1963}
Matheron, G. (1963).
\newblock Principles of {G}eostatistics.
\newblock {\em Economic Geology}, 58(8):1246--1266.

\bibitem[McKay et~al., 1979]{mckay:beckman:conover:1979}
McKay, M., Beckman, R., and Conover, W. (1979).
\newblock A comparison of three methods for selecting values of input variables
  in the analysis of output from a computer code.
\newblock {\em Technometrics}, 21(2):239--245.

\bibitem[Mease and Bingham, 2006]{mease:bingham:2006}
Mease, D. and Bingham, D. (2006).
\newblock Latin {H}yperrectangle {S}ampling for {C}omputer {E}xperiments.
\newblock {\em Technometrics}, 48(4):467--477.

\bibitem[Morris and Mitchell, 1995]{morris:mitchell:1995}
Morris, M.~D. and Mitchell, T.~J. (1995).
\newblock Exploratory designs for computational experiments.
\newblock {\em Journal of Statistical Planning and Inference}, 43:381--402.

\bibitem[Roberts and Rosenthal, 2006]{roberts:rosenthal:2006}
Roberts, G.~O. and Rosenthal, J.~S. (2006).
\newblock Harris recurrence of {M}etropolis-within-{G}ibbs and
  trans-dimensional {M}arkov chains.
\newblock {\em Annals of Applied Probability}, 16(4):2123--2139.

\bibitem[Sacks et~al., 1989a]{sacks:etal:1989:2}
Sacks, J., Schiller, S., Mitchell, T., and Wynn, H. (1989a).
\newblock Design and analysis of computer experiments (with discussion).
\newblock {\em Statistical Science}, 4(4):409--435.

\bibitem[Sacks et~al., 1989b]{sacks:etal:1989:1}
Sacks, J., Schiller, S.~B., and Welch, W.~J. (1989b).
\newblock Designs for computer experiments.
\newblock {\em Technometrics}, 31(1):41--47.

\bibitem[Santner et~al., 2003]{santner:williams:notz:2003}
Santner, T.~J., Williams, B.~J., and Notz, W.~I. (2003).
\newblock {\em The {D}esign and {A}nalysis of {C}omputer {E}xperiments}.
\newblock Springer Series in Statistics. Springer-Verlag, New York.

\bibitem[Schaback, 1995]{schaback:1995}
Schaback, R. (1995).
\newblock Error estimates and condition numbers for radial basis function
  interpolation.
\newblock {\em Advances in Computational Mathematics}, 3(3):251--264.

\bibitem[Schaback, 2007]{schaback:2007}
Schaback, R. (2007).
\newblock Kernel-based meshless methods.
\newblock Technical report, Institute for Numerical and Applied Mathematics,
  Georg-August-University Goettingen.

\bibitem[Stein, 1999]{stein:1999}
Stein, M.~L. (1999).
\newblock {\em Interpolation of spatial data. Some theory for {K}riging}.
\newblock Springer Series in Statistics. Springer-Verlag, New York.

\bibitem[Stein, 2002]{stein:2002}
Stein, M.~L. (2002).
\newblock The screening effect in {K}riging.
\newblock {\em The Annals of Statistics}, 30(1):298--323.

\bibitem[Stinstra et~al., 2003]{stinstra:denhertog:stehouwer:vetjens:2003}
Stinstra, E., den Hertog, D., Stehouwer, P., and Vestjens, A. (2003).
\newblock Constrained maximin designs for computer experiments.
\newblock {\em Technometrics}, 45(4):340--346.

\bibitem[van Dam et~al., 2007]{vandam:husslage:denhertog:melissen:2007}
van Dam, E.~R., Husslage, B., den Hertog, D., and Melissen, H. (2007).
\newblock Maximin {L}atin {H}ypercube {D}esigns in {T}wo {D}imensions.
\newblock {\em Operations Research}, 55(1):158--169.

\end{thebibliography}

\end{document}